\documentclass{PoS}

\usepackage{epsfig}
\usepackage{psfrag}

\title{Regge-like quark-antiquark excitations in the effective-action formalism}

\ShortTitle{Regge-like quark-antiquark excitations in the effective-action formalism}

\author{\speaker{Dmitri Antonov}\\
       Departamento de F\'isica and Centro de F\'isica das Interac\c{c}\~oes Fundamentais,\\ 
       Instituto Superior T\'ecnico, UT Lisboa,\\
       Av. Rovisco Pais, 1049-001 Lisboa, Portugal\\
       E-mail: \email{dima@cfif.ist.utl.pt}}

\author{Jos\'e Em\'ilio F.T. Ribeiro\\
       Departamento de F\'isica and Centro de F\'isica das Interac\c{c}\~oes Fundamentais,\\ 
       Instituto Superior T\'ecnico, UT Lisboa,\\
       Av. Rovisco Pais, 1049-001 Lisboa, Portugal\\
       E-mail: \email{EmilioRibeiro@netcabo.pt}}

\abstract{
Radial excitations of the quark-antiquark string sweeping the Wilson-loop area are considered 
in the framework of the effective-action formalism. Identifying these excitations with the 
daughter Regge trajectories, we find corrections which they produce to the constituent quark mass.
The energy of the quark-antiquark pair turns out to be mostly saturated by the 
constituent quark masses, rather  
than by the elongation of the quark-antiquark string. Specifically, while the constituent quark 
mass turns out to increase as the square root of the radial-excitation 
quantum number, the energy of the string increases only as the fourth root of that number.}

\FullConference{International Workshop on QCD Green's Functions, Confinement and Phenomenology\\
                 5-9 September 2011\\
                 Trento, Italy}

\begin{document}

\section{Introduction}

Low-energy QCD can be characterized by four nonperturbative quantities, of which the gluon condensate
$\langle(gF_{\mu\nu}^a)^2\rangle$ and the vacuum correlation length $\lambda$ (that is, the 
distance at which the two-point, gauge-invariant, correlation function of gluonic field strengths exponentially falls off) are 
related to confinement (so that the string tension corresponding to the two static sources in the fundamental 
representation is $\sigma\propto \lambda^2\langle(gF_{\mu\nu}^a)^2\rangle$~\cite{rev}), while the quark 
condensate $\langle\bar\psi\psi\rangle$, together with the constituent quark mass $m$, characterizes spontaneous 
breaking of chiral symmetry. 
The quark condensate is expressible in terms of gluonic degrees of freedom, 
$\langle\bar\psi\psi\rangle=-\partial\langle\Gamma[A_\mu^a,m]\rangle/\partial m$,
where $\langle\Gamma[A_\mu^a,m]\rangle$, the one-loop effective action, can be represented 
as a world-line integral over the closed quark's trajectories $\{ z_\mu(\tau)\}$ and their 
anticommuting counterparts $\{\psi_\mu(\tau)\}$~\cite{WL},
$$
\langle\Gamma[A_\mu^a,m]\rangle=-2N_{\rm f}
\int_0^\infty
\frac{ds}{s}{\,}{\rm e}^{-m^2s}
\int_{P}^{}{\cal D}z_\mu \int_A^{}{\cal D}\psi_\mu{\,}
{\rm e}^{-\int_0^s d\tau\left(\frac14\dot z_\mu^2+\frac12\psi_\mu\dot\psi_\mu\right)}\times
$$
\begin{equation}
\label{v2}
\times \left\{\biggl<{\rm tr}{\,}{\cal P}
\exp\left[ig\int_0^s d\tau{\,} T^a \left(A_\mu^a\dot z_\mu-\psi_\mu\psi_\nu F_{\mu\nu}^a\right)
\right]\biggr>-N_c\right\}.
\end{equation}
In (\ref{v2}), $N_{\rm f}$ is the number of light-quark flavors, 
$s$ is the Schwinger proper time ``needed'' for the quark to orbit its Euclidean trajectory,
$F_{\mu\nu}^a=\partial_\mu A_\nu^a-\partial_\nu A_\mu^a+gf^{abc}A_\mu^b A_\nu^c$ is the Yang--Mills 
field-strength tensor, and $T^a$'s are the generators of the group SU($N_c$) in the fundamental representation, 
obeying the commutation relation $[T^a,T^b]=if^{abc}T^c$. Notice that, since the quark condensation is generally 
argued to occur due to the gauge fields, the free part of the effective action in Eq.~(\ref{v2}) has been 
subtracted, so that 
$\langle\Gamma[0,m]\rangle=0$. Furthermore, $P$ and $A$ stand there for the periodic ($\int_P^{}\equiv\int_{z_\mu(s)=z_\mu(0)}^{}$) 
and the antiperiodic 
($\int_A^{}\equiv\int_{\psi_\mu(s)=-\psi_\mu(0)}^{}$) boundary conditions, which are imposed, respectively, on the trajectories 
$z_\mu(\tau)$ and their 
Grassmannian counterparts $\psi_\mu(\tau)$ describing $\gamma$-matrices ordered along the trajectory.
The  trajectories obey the condition $\int_0^s d\tau z_\mu(\tau)=0$, meaning that the 
center of each trajectory is the origin. That is, the factor of volume associated with the translation of 
a trajectory as a whole is divided out, and the vector-function $z_\mu(\tau)$ describes only the shape of a closed trajectory,
not its position in space. Finally, throughout this talk, we mean by the quark mass the minimal value of the mass parameter $m$, entering  
Eq.~(\ref{v2}), which renders a finite $\langle\bar\psi\psi\rangle$ --- see Ref.~\cite{3}.

From the mathematical viewpoint, the effective action~(\ref{v2}) represents 
an integral over closed quark trajectories, with the minimal surfaces bounded by those trajectories appearing 
as arguments of the Wilson loops. That is,
$$
\bigl<\Gamma[A_\mu^a,m]\bigr>= -2N_{\rm f}\int_0^\infty\frac{ds}{s}{\rm e}^{-m^2s}
\int_{P}^{}{\cal D}z_\mu \int_{A}^{}
{\cal D}\psi_\mu {\rm e}^{-\int_0^s d\tau\left(\frac14\dot z_\mu^2+\frac12\psi_\mu\dot\psi_\mu\right)}\times
$$
\begin{equation}
\label{effA}
\times \left\{\exp\left[-2\int_0^s d\tau\psi_\mu\psi_\nu\frac{\delta}{\delta s_{\mu\nu}(z(\tau))}\right]
\bigl<W[z_\mu]\bigr>-N_c\right\},
\end{equation}
with the Wilson loop given by
\begin{equation}
\label{st3}
\bigl<W[z_\mu]\bigr>=\left<{\rm tr}{\,}{\cal P}{\,}\exp\left(ig\int_0^s d\tau T^aA_\mu^a\dot z_\mu\right)\right>
\end{equation}
and $\delta /\delta s_{\mu\nu}$ being the area derivative operator, which allows us to recover the spin term  
$\sim\psi_\mu\psi_\nu F_{\mu\nu}^a$ in Eq.~(\ref{v2}). Equation~(\ref{effA}) allows us to reduce
the gauge-field dependence of Eq.~(\ref{v2}) to that 
of a Wilson loop. The Wilson loop is unambiguously defined by the minimal-area surface bounded by the 
contour $z_\mu(\tau)$.

World-line integrals of this type
were first calculated by imposing for the minimal surface a specific parametrization, which in 3D corresponds
to a rotating rod of a variable length~\cite{ps}. We notice that similar parametrizations 
of minimal surfaces spanned by straight-line strings (which interconnect quarks or gluons in the 
one-loop diagrams) are widely used in the existing literature~\cite{dks}. With the 4D parametrization of 
this kind, we were able to obtain, 
within the effective-action formalism, a realistic lower bound for the constituent 
quark mass: $m=460{\,}{\rm MeV}$ for $\langle\bar\psi\psi\rangle\simeq
-(250{\,}{\rm MeV})^3$ (cf. Ref.~\cite{3}). 
For bookkeeping purposes, an explicit derivation 
of this parametrization from the Nambu--Goto string action will be provided in Section~II of this talk.

It happens that for sufficiently large $m$'s, the mean size of the trajectory is smaller than 
the vacuum correlation length $\lambda$, 
so that the nonperturbative 
Yang--Mills fields inside the trajectory can be treated as constant, leading to the area-squared law 
for the Wilson loop~\cite{ds}. By using the world-line representation~(\ref{v2}), 
the area-squared law can be shown (see Ref.~\cite{3} for details) to yield 
the known 
heavy-quark condensate of QCD sum rules, $\langle\bar\psi\psi\rangle_{\rm heavy}\propto -\langle(gF_{\mu\nu}^a)^2\rangle/m$,
in which case $m$ becomes the constituent mass of a heavy quark. 
For lighter quarks, 
the mean size of their Euclidean trajectories exceeds $\lambda$, 
so that one can expect the area-squared law to be 
morphed into an area law. However, unlike static color sources transforming according to the fundamental 
representation of SU($N_c$), which yield an area law with a constant string tension $\sigma$, light quarks are subject 
to a zigzag-type motion. 
This type of motion can only be reconciled with the area law for some 
effective scale-dependent string tension $\tilde\sigma(s)\propto 1/s$, so as to obtain a nonvanishing 
$\langle\bar\psi\psi\rangle$~\cite{3}. We notice that the 
fractalization of quark trajectories, which takes place
upon the deviation from the heavy-quark limit, has been studied
in Ref.~\cite{99} in terms of the velocity-dependent quark-antiquark 
potentials. 

It is then clear from the above discussion that the minimal value for the mass parameter $m$  and the 
mean size of the quark trajectory are not independent 
from each other so that, in general, $\bigl<\Gamma[A_\mu^a,m]\bigr>$ is a functional of the 
minimal area. To this we have to add that 
the quark condensate characterizes the QCD vacuum, 
and, therefore, should remain invariant under the variations of the minimal area. 
The final ingredient should come from energy conservation: the total energy  of a given excited quark-antiquark system should be equal to the sum of the 
quark masses and the energy stored in the quark-antiquark string.
Then, the purpose of our analysis is to study the effective action as a functional of the minimal area 
for quark-antiquark excited systems, that is, 
to evaluate the corresponding variation 
$\delta m$ such that $\langle\bar\psi\psi\rangle$ remains constant for a given series of quark-antiquark radial excitations. 
At this point, we need a physical input 
on how to quantify these excitations. A natural choice will be to link these excitations to the 
daughter Regge trajectories~\cite{rev1}. In this work, we adopt this point of view.

The talk is organized as follows. In the next Section, using as an input the excitation energy corresponding 
to the $n$-th daughter Regge trajectory,
we introduce an Ansatz for the scaling 
factor $Z_n$, which describes an increase in the area of the excited-string world sheet. 
In this way, we consider only the radial excitation modes of the quark-antiquark pair, which 
correspond to 
the ``breathing modes'' of the string world sheet. Consideration of angular excitations, which would correspond to the disclinations of 
the world sheet, lies outside the scope of our present analysis. 
Then, we proceed to a selfconsistent determination of the 
``critical index'' $\gamma$, which defines that Ansatz. With this knowledge at hand, we calculate
the constituent quark-mass correction $\delta m_n$ as a function of the radial-excitation quantum number $n$.
In Section~III, using the 
large-$n$ asymptotes of the formulae obtained for $Z_n$ and $\delta m_n$, we  
show that the primary contribution to the excitation energy of the quark-antiquark pair stems from the constituent 
quark mass $m_n$, and not from the area increase. We also obtain the lowest (that is, $n=1$) correction 
to the constituent quark mass, $\delta m_1$, which turns out to be about 26~MeV. Also in Section~III, 
we present some concluding remarks and an outlook.

\section{A correction to the constituent quark mass}

In this Section, we derive a general expression for the correction to the 
constituent quark mass, coming from radial excitations of the quark-antiquark string sweeping the surface of the Wilson loop.
To this end, we use for the eigenenergies of radial excitations of the quark-antiquark pair 
the Regge formula~\cite{rev1}: 
\begin{equation}
\label{0m}
E_n=\sqrt{\pi\sigma (4n+3)}.
\end{equation} 
Here, $n$ is the quantum number of a radial excitation, and  
$\sigma\simeq(440{\,}{\rm MeV})^2$ stands for the string tension in  
the fundamental representation of the group SU(3).

The energy gap $E_n-E_0$ can be filled in by both deformations of the quark-antiquark string
and/or by increasing in the quark constituent masses. That is, 
\begin{equation}
\label{1}
E_n-E_0=\sigma\cdot(L_n-L_0)+2(m_n-m_0),
\end{equation}
where $L_n$'s are the eigenvalues of the length of the string. Notice that 
$m_n$ turns out to be an implicit function of $L_n$.
We furthermore denote by $2R_n$ the diameter of the semiclassical Euclidean trajectory performed by the 
quark in the $n$-th excited state of the system. The value of 
$2R_n$ cannot exceed some $2R_{n,{\,}{\rm max}}$, at which string-breaking occurs. 
The string elongation is given by the ratio $L_n/L_0$, where in the $(n=0)$ state we must have $2R_0=L_0$.

Since we will be calculating the constituent quark masses $m_n$ in the units of $\sqrt{\sigma}$, it is  
convenient to introduce a dimensionless function 
\begin{equation}
\label{fN}
f_n\equiv m_n/\sqrt{\pi\sigma}.
\end{equation}
The sought correction to the constituent quark mass
can be obtained from the effective action~(\ref{v2}) if one uses there for the surface,  
entering the Wilson loop, the world sheet of the excited quark-antiquark string. To this end, let us start by 
defining the scaling factor $Z_n$ which describes an enlargement of the world-sheet area
over its value in the $(n=0)$ state. We have 
\begin{equation}
\label{0}
\frac{L_n}{L_0}=1+S_n,~~~~ {\rm where}~~~~ 
S_n\equiv\frac{\sqrt{\pi/\sigma}}{2R_0}\left[\sqrt{4n+3}-\sqrt{3}-2\cdot\delta f_n\right],
\end{equation}
and $\delta f_n\equiv f_n-f_0$. Accordingly, the scaling factor, which describes an increase of the 
area of the string world sheet in the $n$-th excited state, reads 
\begin{equation}
\label{z1}
Z_n=\left(\frac{L_n}{L_0}\right)^2=(1+S_n)^2.
\end{equation}

We further notice that the role of a global characteristic of the quark trajectory can be played either 
by its semiclassical radius $R_n$, or by the proper time $s$ during which the trajectory is orbited by the quark.
Therefore, these two quantities are related to each other through a scaling relation of the 
form
\begin{equation}
\label{rad}
R_n=R_{n,{\,}{\rm max}}\cdot\left(\frac{s}{s_{\rm max}}\right)^\gamma .
\end{equation} 
Here $s_{\rm max}$ 
is the proper time necessary for the quark to orbit a trajectory of the maximum diameter
$2R_{n,{\,}{\rm max}}$ equal to the string-breaking distance. The actual value of the ``critical index'' $\gamma$ will be selfconsistently determined below.
Thus, we get for the correcting term $S_n$ in Eq.~(\ref{z1}) the 
following expression: 
\begin{equation}
\label{k78}
S_n=S_n(s)=\frac{\xi_n}{s^{\gamma}},~~~ {\rm where}~~~ 
\xi_n\equiv\frac{\sqrt{\pi/\sigma}{\,} s_{\rm max}^\gamma}{2 
R_{0,{\,}{\rm max}}}\left[\sqrt{4n+3}-\sqrt{3}-2\cdot\delta f_n\right].
\end{equation}

We proceed now to the discussion of parametrizations for the minimal area of the string world sheet and for the 
Wilson loop. For the minimal area of the unexcited-string world sheet we use   
the following parametrization: 
\begin{equation}
\label{s4d}
S_{\rm 4D}=\frac{1}{2\sqrt{2}}\int_0^s d\tau|\varepsilon_{\mu\nu\lambda\rho}z_\lambda\dot z_\rho|.
\end{equation}
It represents a four-dimensional generalization of $S_{\rm 3D}=\frac12
\int_0^s d\tau|{\bf z}\times\dot{\bf z}|$ (cf. Ref.~\cite{ps}), 
which is the area-functional of a surface swept out by a rotating rod of a variable length. 
We notice that $S_{\rm 4D}$ stems directly from the usual formula for the 
area (corresponding to the Nambu--Goto string action) upon the parametrization of the surface by the vector-function 
$w_\mu(\zeta_1,\zeta_2)=\zeta_2\cdot z_\mu(\zeta_1/\sigma)$, where $\zeta_1=\sigma\tau$ and $\zeta_2\in[0,1]$.
Indeed, the usual formula for the area reads ${\cal A}=\int_{0}^{\sigma s} d\zeta_1\int_0^1 d\zeta_2{\,} 
\sqrt{\det g_{ab}}$, where $g_{ab}=\partial_a w_\mu\cdot
\partial_b w_\mu$
is the induced-metric tensor, and each of the indices $a$ and $b$ takes the values 1 and 2.
Using the above parametrization for $w_\mu(\zeta_1,\zeta_2)$, 
one can then readily prove the following equality: 
$$S_{\rm 4D}=
{\cal A}=\frac{1}{2}\int_0^s d\tau\sqrt{
z_\mu^2\dot z_\nu^2-(z_\mu\dot z_\mu)^2}.$$
Next, we follow Ref.~\cite{3} for what concerns the parametrization of the Wilson loop~(\ref{st3}), which can be 
written in the form
\begin{equation}
\label{mW}
\langle W[z_\mu]\rangle=
\frac{N_c}{2^{\alpha-1}\Gamma(\alpha)}\cdot(\tilde\sigma|\Sigma_{\mu\nu}|)^\alpha\cdot
K_\alpha(\tilde\sigma|\Sigma_{\mu\nu}|).
\end{equation}
Here, $\Sigma_{\mu\nu}=\varepsilon_{\mu\nu\lambda\rho}\int_0^s d\tau z_\lambda\dot z_\rho$ is the integrated 
surface element, whose absolute value is implied in the sense 
that $|\Sigma_{\mu\nu}|=\Bigl(2\sum\limits_{\mu<\nu}^{}\Sigma_{\mu\nu}^2\Bigr)^{1/2}$. Furthermore, 
$\Gamma(x)$ and $K_\alpha(x)$ in Eq.~(\ref{mW}) stand, respectively, for Gamma- and MacDonald functions, 
$\alpha\gtrsim 1$ is some parameter, and 
$\tilde\sigma=\tilde\sigma(s)$ stands for the effective string tension of dynamical quarks. 
As it was shown in Ref.~\cite{3}, Eq.~(\ref{mW}) provides an interpolation between 
the area law for large loops and the area-squared law~\cite{ds} for small loops. 
This statement is illustrated by Figs.~1 and 2, which compare the combined parametrization~(\ref{mW}), 
for $\alpha=1.9$, with the area and the area-squared laws. For this illustration, we choose a circular 
contour with the minimal area $S$, set $N_c=3$, and use the relation~\cite{ds} 
$\langle(gF_{\mu\nu}^a)^2\rangle=\frac{72}{\pi}{\,}\frac{\sigma}{\lambda^2}$, where 
$\lambda=1.72{\,}{\rm GeV}^{-1}$~\cite{wq} for the case of QCD with dynamical quarks considered here. 
The chosen value of $\alpha=1.9$ has been shown in Ref.~\cite{3} to provide the best analytic 
approximation of the area-squared law by the combined parametrization~(\ref{mW}) at small distances. 
Figure~1 illustrates the efficiency of this approximation.

\begin{figure}
\psfrag{x}{$\sigma S$}
\epsfig{file=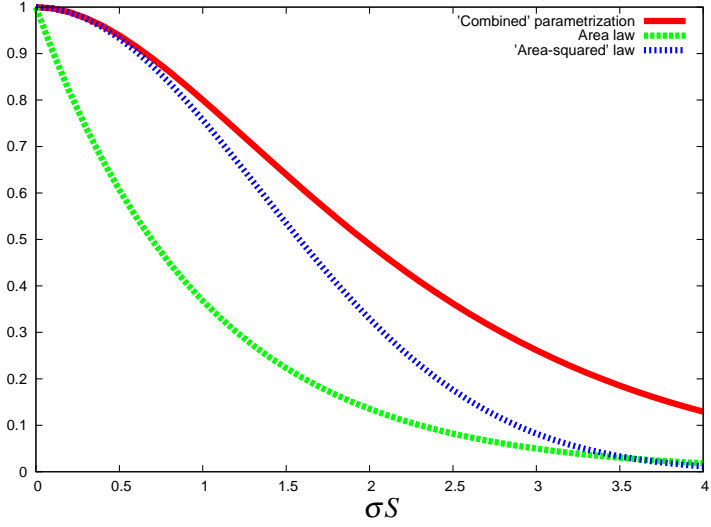, width=100mm}
\caption{$\frac{1}{2^{\alpha-1}\Gamma(\alpha)}\cdot (\sigma S)^\alpha\cdot K_\alpha(\sigma S)$ at $\alpha=1.9$,~~ ${\rm e}^{-\sigma S}$,~~  ${\rm e}^{-\frac{\langle(gF_{\mu\nu}^a)^2\rangle}{48N_c}{\,}S^2}$. 
The value $\sigma S=4$ 
corresponds to the radius of the contour equal to 0.51~fm.}
\end{figure}

\begin{figure}
\psfrag{x}{$\sigma_{\rm f}S$}
\epsfig{file=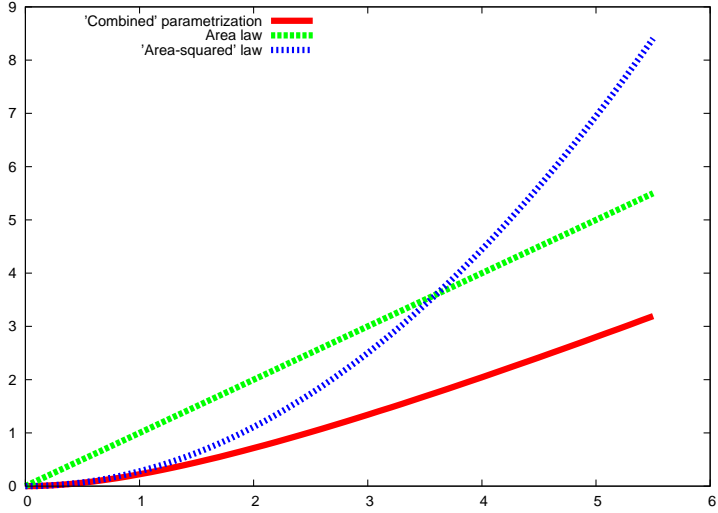, width=100mm}
\caption{$-\ln\left[\frac{1}{2^{\alpha-1}\Gamma(\alpha)}\cdot 
(\sigma S)^\alpha\cdot K_\alpha(\sigma S)\right]$ at $\alpha=1.9$,~~ $\sigma S$,~~  
$\frac{\langle(gF_{\mu\nu}^a)^2\rangle}{48N_c}{\,}S^2$. The value $\sigma S=5.5$ 
corresponds to the radius of the contour equal to 0.6~fm. The gap between $-\ln[\cdots]$ and $\sigma S$, being 
$\sim\ln(\sigma S)$ at large distances, is irrelevant for the static potential.}
\end{figure}

We are now in a position to calculate $\langle\Gamma[A_\mu^a,m_n]\rangle$, which is defined by Eq.~(\ref{v2})
with $m$ replaced by $m_n$. For this purpose, we 
multiply the infinitesimal surface element  
$d\tau z_\lambda\dot z_\rho$ in $\Sigma_{\mu\nu}$ above by the 
scaling factor $Z_n$. 
Following Ref.~\cite{3}, we arrive at the expression 
\begin{equation}
\label{nr}
\langle\Gamma[A_\mu^a,m_n]\rangle=-2N_{\rm f}N_c\int_0^\infty\frac{ds}{s}{\,}{\rm e}^{-m_n^2s}\cdot
\frac{\alpha(\alpha+1)(\alpha+2)}{(2\pi\tilde\sigma^2)^3}\left(\prod\limits_{\mu<\nu}^{}\int_{-\infty}^{+\infty}
dB_{\mu\nu}\right)\frac{
I(Z_n,{\cal F}_{\mu\nu})}{\left(1+\frac{B_{\mu\nu}^2}{4\tilde\sigma^2}\right)^{\alpha+3}},
\end{equation}
where
\begin{equation}
\label{33}
I(Z_n,{\cal F}_{\mu\nu})=\int_{P}^{}{\cal D}z_\mu \int_{A}^{}{\cal D}\psi_\mu
{\rm e}^{-\int_0^s d\tau\left(
\frac14\dot z_\mu^2+\frac12\psi_\mu\dot\psi_\mu+\frac{i}{2}Z_n{\,}{\cal F}_{\mu\nu}z_\mu\dot z_\nu-
iZ_n{\,}{\cal F}_{\mu\nu}\psi_\mu\psi_\nu\right)}
\end{equation}
and ${\cal F}_{\mu\nu}\equiv\varepsilon_{\mu\nu\lambda\rho}B_{\lambda\rho}$. 

The world-line integral~(\ref{33}) describes an 
infinite sum of one-loop quark diagrams, each having its own number of external lines of the 
auxiliary gauge field $B_{\mu\nu}$. Similarly to what was done in Ref.~\cite{3}, we retain in this diagrammatic expansion only the two leading 
terms -- the one corresponding to a free quark, which cancels out in Eq.~(\ref{v2}), and the one corresponding to a diagram 
with two external lines of the gauge field.  The latter yields the following expression:
\begin{equation}
\label{kn8}
I(Z_n,{\cal F}_{\mu\nu}) =\frac{1}{(4\pi s)^2}\left[\frac{4}{3}(s B Z_n)^2+ 
{\cal O}
\bigl((s B Z_n)^4\bigr)\right],
\end{equation}
where $B=\Bigl(\sum\limits_{\mu<\nu}^{}B_{\mu\nu}^2\Bigr)^{1/2}$.

We start our analysis with the 
limit $n\gg 1$, where one can approximate the factor $Z_n$ in Eq.~(\ref{kn8}) by $S_n^2$. 
In the same large-$n$ limit, it is legitimate to 
disregard the quartic and the higher terms in the expansion~(\ref{kn8}), provided 
the amplitude $B$ is bounded from above as 
\begin{equation}
\label{kn9}
B<\frac{1}{sS_n^2}=\frac{s^{2\gamma-1}}{\xi_n^2}.
\end{equation}
In the last equality, we have used the explicit parametrization~(\ref{k78}) of $S_n$ in terms of $s$.
Now, as it was already mentioned in Introduction,
the quark condensate, $\langle\bar\psi\psi\rangle=-\frac{\partial}{\partial m_n}{\,}
\langle\Gamma[A_\mu^a,m_n]\rangle$, being the quantity which characterizes the QCD vacuum, should remain $n$-independent. 
For this to be possible, the 
energy of string excitations should be largely absorbed by the constituent quark masses $m_n$, and 
it will be demonstrated below that this is what indeed happens. 

An expression for the quark condensate following from Eqs.~(\ref{nr})-(\ref{kn9}) reads
\begin{equation}
\label{444}
\langle\bar\psi\psi\rangle=-\frac{\alpha(\alpha+1)(\alpha+2)N_{\rm f}}{8\pi^2}\cdot m_n\int_0^\infty ds{\,}
\frac{{\rm e}^{-m_n^2s}}{\tilde\sigma^6}{\,}S_n^4\int_0^{s^{2\gamma-1}/\xi_n^2}dB{\,}\frac{B^7}{\left(1+
\frac{B^2}{2\tilde\sigma^2}\right)^{\alpha+3}},
\end{equation}
where from now on we set $N_c=3$.
The $B$-integration in this formula can be performed analytically, to yield 
\begin{equation}
\label{jj}
\langle\bar\psi\psi\rangle=-\frac{3N_{\rm f}}{4\pi^2}\cdot m_n
\int_0^\infty ds
{\,}{\rm e}^{-m_n^2s}\cdot\frac{\tilde f[A_n(s),\alpha]}{2s^2A_n(s)},
\end{equation}
where
\begin{equation}
\label{aN8}
A_n(s)\equiv\frac12\left(\frac{s^{2\gamma-1}}{\tilde\sigma\xi_n^2}\right)^2,
\end{equation}
while a somewhat complicated function 
$\tilde f[A,\alpha]$
was introduced in Ref.~\cite{3}. An important property of this function is that, for $\alpha\gtrsim 1$ of interest, the ratio $\tilde f[A,\alpha]/A$, as a function of $A$, has a maximum at $A\ll 1$, which sharpens and reaches a finite value ($\simeq 1.18$), with the further increase of $\alpha$.

In order for the quark condensate to stay finite 
in the small-mass limit, one should be able to represent $\frac{\tilde f[A_n(s),\alpha]}{2s^2A_n(s)}$ 
in the form (cf. Refs.~\cite{4,3})
\begin{equation}
\label{87}
\frac{\tilde f[A_n(s),\alpha]}{2s^2A_n(s)}=
\frac{\sigma_0^{3/2}}{\sqrt{s}},
\end{equation}
where $\sigma_0$ is some parameter of dimensionality (mass)$^2$ unambiguously related to the 
phenomenological value of $\langle\bar\psi\psi\rangle$. 
Moreover, this representation should remain valid  
up to the values of the proper time such that 
\begin{equation}
\label{bo65}
m_n^2{\,}s_{\rm max}\gtrsim 1.
\end{equation}
Equation~(\ref{87})
can equivalently be written as
\begin{equation}
\label{5g}
\frac{\tilde f[A_n,\alpha]}{A_n}=x,~~~ {\rm where}~~~  
x\equiv2(\sigma_0 s)^{3/2}.
\end{equation}
Owing to the above-mentioned form of the function $\tilde f[A,\alpha]/A$, a
solution to Eq.~(\ref{5g}), which 
provides a physical decrease of $\tilde\sigma$ with $s$, reads
$A_n\simeq x^\varepsilon$, where $\varepsilon\to 0$ for $\alpha\gtrsim 1$ of interest, and 
$x\lesssim 1$ (cf. Ref.~\cite{3}).
Therefore, to a very good approximation, one can set in Eq.~(\ref{aN8}) $A_n \simeq 1$. 
This allows us to obtain the 
actual value of the power $\gamma$ in the initial Ansatz~(\ref{rad}).
To this end, we notice that, due to the energy conservation in the quark-antiquark system, a variation of the semiclassical 
radius $R_n$ of the 
trajectory leads to the variation $\delta m_n = \tilde\sigma{\,}\delta R_n$ of the constituent quark mass.
Therefore, a difference between the values of the radius $R_n$ in the $n$-th and the 0-th states, $\delta R_n\equiv R_n-R_0$, 
reads
\begin{equation}
\label{Rx1}
\delta R_n=\frac{\delta m_n}{\tilde\sigma}=
\delta m_n\cdot\frac{\sqrt{2}\xi_n^2}{s^{2\gamma-1}},
\end{equation}
where Eq.~(\ref{aN8}) has been used at the final step. 
Now, in order for Eqs.~(\ref{rad}) and (\ref{Rx1}) to have the same $s$-dependence, $\gamma=1/3.$ 

With this value of $\gamma$ at hand, we can now 
calculate the correction $\delta m_n$ to the constituent quark mass, which is produced by the radial 
excitations of the quark-antiquark string. To this end, we 
first insert the value of $\gamma=1/3$ into Eq.~(\ref{Rx1}), that leads to the following 
relation:
$(\delta R_n)_{\rm max}=\sqrt{2}{\,}\delta m_n\cdot\xi_{n,{\,}{\rm max}}^2{\,} s_{\rm max}^{1/3}$.
Furthermore, for $\xi_{n,{\,}{\rm max}}$ in this formula we use its expression provided by 
Eq.~(\ref{k78}) with $\gamma=1/3$. That yields
\begin{equation}
\label{rm7}
(\delta R_n)_{\rm max}=\pi\sqrt{2}\cdot\frac{\delta m_n{\,} s_{\rm max}}{\sigma R_{0,{\,}{\rm max}}^2}
\left[\sqrt{n+\frac34}-\sqrt{\frac34}-(\delta f_n)_{\rm min}\right]^2.
\end{equation}
Next, we use the approximation $\delta m_n{\,} s_{\rm max}\simeq 1/(\delta m_n)_{\rm min}$,
which reflects the fact that the trajectory of a maximum size (and therefore requiring the maximum proper time 
to be orbited) is reached when the value of the 
constituent quark mass is minimal. [Notice that this approximation parallels  
condition~(\ref{bo65}).] Substituting this approximation into Eq.~(\ref{rm7}), we arrive at the 
following equation:
\begin{equation}
\label{mMin}
(\delta m_n)_{\rm min}\simeq
\frac{\pi\sqrt{2}}{\sigma R_{0,{\,}{\rm max}}^2
(\delta R_n)_{\rm max}}\left[\sqrt{n+\frac34}-\sqrt{\frac34}
-(\delta f_n)_{\rm min}\right]^2.
\end{equation}
In order to solve this equation, we represent it entirely in terms of $(\delta f_n)_{\rm min}$. That can 
be done by virtue of the relation $(\delta m_n)_{\rm min}=\sqrt{\pi\sigma}{\,}(\delta f_n)_{\rm min}$, which
stems from Eq.~(\ref{fN}). As a result, we obtain the following quadratic equation:
\begin{equation}
\label{q23}
(\delta f_n)_{\rm min}+b\cdot (\delta f_n)_{\rm min}^{1/2}-\left(\sqrt{n+\frac34}-\sqrt{\frac34}\right)=0,~~~ 
{\rm where}~~~ b\equiv\frac{\sigma^{3/4} 
R_{0,{\,}{\rm max}}(\delta R_n)_{\rm max}^{1/2}}{(2\pi)^{1/4}}.
\end{equation}
A solution to this equation yields the sought 
correction to the constituent quark mass:
\begin{equation}
\label{mN}
(\delta m_n)_{\rm min}=\sqrt{\pi\sigma}\cdot(\delta f_n)_{\rm min}=
\frac{\sqrt{\pi\sigma}}{4}\left[\sqrt{b^2+4\left(\sqrt{n+\frac34}-\sqrt{\frac34}\right)}-b
\right]^2.
\end{equation}
The limits of this formula at large and small $n$'s will be analyzed in the next Section.

\section{The limiting cases of large and small excitations. Concluding remarks}

Ignoring for a moment the effect of string-breaking, we see that the obtained
$(\delta m_n)_{\rm min}$, Eq. (\ref{mN}), vanishes in the limit of  
$b\to\infty$, which corresponds to 
the quark-antiquark string of an infinite length [cf. the definition of $b$ in Eq.~(\ref{q23})].
In reality, however, the string-breaking phenomenon imposes an upper limit on the possible values of 
$b$. Indeed, the upper limit for  
both $R_{0,{\,}{\rm max}}$ and $(\delta R_n)_{\rm max}$ is given by $d_{\rm s.b.}/2$, where 
$d_{\rm s.b.}$ is the string-breaking distance. Lattice simulations and analytic studies~\cite{sb} 
suggest for this distance the value of 
$d_{\rm s.b.}\simeq1.5{\,}{\rm fm}$. Using also the phenomenological 
value of $\sigma\simeq(440{\,}{\rm MeV})^2$, we get $b\lesssim1.37$.
Therefore, we find from Eq.~(\ref{mN}) the realistic asymptotic behavior of the constituent quark mass to be
\begin{equation}
\label{mm8}
(\delta m_n)_{\rm min}\to\sqrt{\pi\sigma n}~~~ {\rm for}~~~ n\gg 1.
\end{equation}
Comparison of this result 
with the initial Eqs.~(\ref{0m}) and (\ref{1}) shows that, in the large-$n$ limit, the leading 
contribution to the excitation energy $E_n$ of the quark-antiquark pair 
stems from the constituent quark masses. 
Indeed, one can perform the large-$n$ expansion of $(\delta f_n)_{\rm min}$ given by Eq.~(\ref{mN}), 
which yields
$(\delta f_n)_{\rm min}=\sqrt{n}\left[1-bn^{-1/4}+{\cal O}(n^{-1/2})\right]$. Then, 
inserting this expansion into the formula for $S_n$,
Eq.~(\ref{k78}), we obtain the leading large-$n$ behavior 
\begin{equation}
\label{s9}
S_n\bigr|_{s=s_{\rm max}}\to\frac{\sqrt{\pi/\sigma}}{R_{0,{\,}{\rm max}}}\cdot b\;n^{1/4},
\end{equation}
which is subdominant compared to Eq.~(\ref{mm8}).
Recalling Eq.~(\ref{0}), we 
conclude that 
$$m_{n,{\,}{\rm min}}\sim L_n^2.$$ 
Thus, the constituent quark mass appears as a 
primary ingredient of the excitation energy of the quark-antiquark pair 
in the large-$n$ limit, 
whereas the elongation of the 
string plays only a secondary role. Still, we observe an increase of $S_n$ with $n$, which, 
for sufficiently large $n$'s, 
validates the approximation $Z_n\simeq S_n^2$ used after Eq.~(\ref{kn8}).

Let us now evaluate a correction to the constituent quark mass, which is associated with the 
$(n=1)$ excitation. We note that this excitation is developed just on top of the maximally-stretched 
unexcited-string configuration. For this reason, one can use for 
such an evaluation the above-adopted maximum values of $b=1.37$ and $R_{0,{\,}{\rm max}}=0.75{\,}{\rm fm}$, 
and also set $s=s_{\rm max}$. 
Extrapolating then Eq.~(\ref{s9}) down to $n=1$, we get
$S_1\equiv\left.S_1\right|_{s=s_{\rm max}}=1.45$. The fact that this extrapolation to $n=1$ 
of the initial parametrically large result of Eq.~(\ref{s9}) leads to $S_1\sim 1$, 
signals the need to introduce some correcting numerical factor $k$. It can be defined through the relation
$(kS_1)^2=Z_1$, where again $Z_1=(1+S_1)^2$. This equation yields the value of 
$k=1.69$. Next, according to Eq.~(\ref{s9}), the multiplication 
of $S_1$ by a factor of $k$ is equivalent to the multiplication of $b$ by such a factor. 
This observation yields the corrected value of $b=2.32$.
Inserting it into Eq.~(\ref{mN}), we get the sought estimate for the $(n=1)$ 
correction to the constituent quark mass: 
$$(\delta m_1)_{\rm min}=26.0{\,}{\rm MeV}.$$
This value looks like a reasonable additive correction to the leading result, $m=460{\,}{\rm MeV}$,
quoted in the Introduction.

In conclusion, we notice that excitations of the quark-antiquark pairs can in general lead to an increase
of the constituent quark mass and to an elongation of the quark-antiquark string. In this talk, we have 
shown that, for large radial excitations $n$, the constituent quark mass grows as ${\cal O}(n^{1/2})$,
while the length of the string grows only as ${\cal O}(n^{1/4})$. This result clearly means that 
the excitation energy of a quark-antiquark pair stems mostly from the increase of the 
constituent quark masses and not from string elongation. Thus, at least within the effective-action formalism, 
excited quark-antiquark 
bound states tend to have essentially the same size, irrespective of their radial excitations. This is true even if we had energy 
dependence for bound states different from that of Eq.~(\ref{0m}), because that will be just another prescription 
for the area scaling and hence, the final qualitative result would not depend 
on actual details on how this scaling is obtained, but just from the fact that for a given scaling up of the area, 
the mass would go like the 
square of the string elongation, whereas the string energy would dimensionally go with $\sigma L_n$. Finally, such size 
stiffness will preclude decay channels to become large with string elongations, because they chiefly measure the size of 
the parent hadron and that does not change appreciably. We also emphasize that the adopted calculational method 
does not rely on any specific class of the string deformations 
(such as, e.g., the normal modes). Finally, it looks natural to apply the present approach  
to the description of an interesting lattice result~\cite{kl} that, for quarks in the fundamental representation, 
the deconfinement and the chiral-symmetry-restoration temperatures are nearly the same, whereas for quarks in the 
adjoint representation, the chiral-symmetry-restoration temperature exceeds the 
deconfinement one by a factor of 8. Work in this direction is currently in progress.

\begin{acknowledgments}

\noindent
D.A. thanks the organizers of the QCD-TNT-II International Workshop for an opportunity to present these results 
in a very nice and stimulating atmosphere.
The work of D.A. was supported by the Portuguese Foundation for Science and Technology
(FCT, program Ci\^encia-2008) and by 
the Center for Physics of Fundamental Interactions (CFIF) at Instituto Superior
T\'ecnico (IST), Lisbon. 
\end{acknowledgments}

\end{document}